\documentstyle[preprint,aps]{revtex}

\input epsf

\begin{document}

\preprint{DUKE-TH-96-118}

\title{Study of Chaos and Scaling in Classical SU(2) Gauge Theory}

\author{Berndt M\"uller}

\address{Department of Physics, Duke University,\\
Durham, North Carolina, 27708-0305, USA.}

\maketitle

\date{DRAFT: \today}

\begin{abstract}
Following a recent suggestion by Nielsen, Rugh, and Rugh, we study the 
energy scaling of the maximal Lyapunov exponent of classical Hamiltonian 
SU(2) lattice gauge theory.  It is shown that the conjectured scaling 
behavior $\lambda_0\sim E^{1/4}$ at small energies on the lattice is a 
finite-time artifact.  New numerical results for the maximal Lyapunov 
exponent are presented for lattices up to size $20^3$ and over two orders 
of magnitude in the energy per plaquette.

\end{abstract}

\pacs{??}

\section{Introduction}

Numerical solutions of the classical Yang-Mills equation in Minkowski
space have recently attracted some interest as a nonperturbative tool
for studying the dynamics of gauge fields.  Much of this interest was
motivated by the desire to calculate the rate of baryon number
fluctuations in the high-temperature phase of the electro-weak gauge
theory \cite{Amb91,Amb92,Amb95}.  Other studies have concentrated on
the spectrum of Lyapunov exponents of the Hamiltonian gauge field
dynamics defined on a spatial lattice \cite{Mul92,Gon94,Bir94}.

In a recent manuscript, Nielsen, Rugh, and Rugh \cite{Nie96} have
analyzed the problems associated with the ultraviolet divergence of
classical gauge fields and their implications for the Lyapunov
exponents of these nonlinear dynamical systems.  They showed that,
when the divergences are regulated by a short-distance lattice
cut-off, the high energy behavior of the Lyapunov exponents is
dominated by lattice artifacts.  For low energies they conjectured
that the energy scaling of the maximal Lyapunov exponent $\lambda_0$
is the same as that found in the limit of homogeneous gauge potentials
\cite{Chi81}:  $\lambda_0 \sim E^{1/4}$.  They argued further that the
approximately linear energy dependence observed at intermediate
energies does not have significance in terms of ``continuum physics''.

By contrast, Bir\'o et al. \cite{Bir95} had argued that the linear
energy dependence of the maximal Lyapunov exponent could be related to
the linear temperature dependence of the plasmon damping rate
\cite{Bra90} in the thermal quantum gauge theory.  The apparent
deviations from this scaling behavior were interpreted as numerical
artifacts caused by the loss of reliability of the procedure used in
\cite{Mul92} for the determination of the maximal Lyapunov exponent.

The present study was motivated by the desire to resolve this
controversy by a more careful investigation of lattice artifacts,
combined with a more reliable algorithm for the calculation of
Lyapunov exponents.  As will be seen, this study clearly exhibits the
presence of finite-size and finite-time effects at small energies.  
It shows the need for very long runs on large lattices in order to obtain
accurate results in the low energy limit.

\section{General Considerations}

The numerical studies reported here were performed by integrating the
classical SU(2) gauge fields on a spatial lattice in the Hamiltonian
formulation (in the temporal gauge) with a continuous time
variable.  These equations are covariant against arbitrary
space-dependent gauge transformations, but not against gauge
transformations that depend on time.  The equations of motion for
the link variables $U_\ell$ and the numerical algorithm employed in 
their solution are discussed in detail in \cite{Bir94}.  
We denote the lattice spacing by $a$, the
linear extent of the cubic lattice by $(Na)$, and the gauge coupling 
constant by $g$.  As noted in \cite{Bir94}, $g$ and $a$ can be
scaled out of the Hamiltonian of the classical Yang-Mills theory.  
The combination $g^2Ha$ is dimensionless, and the equations for the 
lattice gauge field can be written in parameter-free form.  However, in
order to facilitate physical arguments we will retain $g$ and $a$ as
parameters throughout our discussion here.

In the original work on this topic \cite{Mul92} the maximal Lyapunov
exponent was obtained by following two initially close gauge field
configurations for a period of time and observing the exponential
growth of an appropriately determined distance between the gauge
fields.  The accuracy of this ``slope method'' is principally limited
because the distance between two gauge fields is bounded from above
due to the compactness of phase space for a fixed total energy.  The
method loses its reliability completely at small energies because the
exponential divergence is then superseded by distance fluctuations.  A
more reliable method for the calculation of the maximal Lyapunov
exponent uses the rescaling algorithm (see \cite{Bir94}, section
3.6).  Here one follows two neighboring field configurations for an
arbitrarily long time, periodically rescaling the distance to a small
value.  The accuracy of the obtained value for $\lambda_0$ here is
only limited by the time one is willing to spend on the calculation.
In the calculation reported here we have used a gauge non-invariant
distance measure
\begin{equation}
D\left[ U_{\ell},E_{\ell}; U'_{\ell},E'_{\ell}\right] = \sum_{\ell}
\left( \vert U_{\ell}-U'_{\ell}\vert^2 + \vert
E_{\ell}-E'_{\ell}\vert^2\right). \label{e1}
\end{equation}
This leads to the same results for the Lyapunov exponent as a gauge
invariant distance measure, because the equations of motion conserve
Gauss' law, hence gauge degrees of freedom do not contribute to the
exponential growth of the distance $D$ \cite{Gon94}.

Before I begin to discuss the new results, it is useful to recall some
peculiarities of the classical gauge field theory defined on a
lattice.  As numerical calculations have shown, the distribution of
the energy density in the gauge field rapidly becomes thermal
\cite{Bir94}.  In the weak coupling limit, the ``temperature''
$T$---i.e. the slope parameter of the energy density distribution---is
simply related to the total energy per elementary plaquette, $E_p =
H/3N^3$,
\begin{equation}
E_p = {2\over 3} (N_c^2 -1)T = 2T \quad (g^2E_pa \ll 1). \label{e2}
\end{equation}
where $N_c$ is the number of colors (here $N_c=2)$.  The energy
density of the gauge field is
\begin{equation}
\varepsilon = 3E_p/a^3 = 6T/a^3. \label{e3}
\end{equation}
This expression, which diverges in the limit $a\to 0$, is
indicative of the fact that most of the energy contained in the 
thermalized classical gauge field resides in short wavelength modes.

The next quantity of interest is the Debye screening length $\mu^{-1}$ 
of the thermal gauge field.  On the lattice one finds in the $N\to\infty$
limit:
\begin{equation}
\mu^2 = 2N_cg^2 \int {d^3k\over (2\pi)^3} \left(
-{\partial\over\partial\omega}\right) n(\omega) \approx 1.22\,N_c\, 
{g^2T\over \pi a},
\label{e4}
\end{equation}
where $\omega=\vert\vec k\vert$ and $n(\omega) = T/\omega$ is the
classical limit of the Bose distribution function.  The combination
$g^2T$ has the dimension of an inverse length.  The Debye screening 
length $\mu^{-1}$ vanishes in the limit $a\to 0$ as 
$\mu^{-1}\propto \sqrt{a}\to 0$.

The length scale $(g^2T)^{-1}$, which is associated with the
correlation length of static magnetic fields in the thermal quantum
gauge theory, is a classical length scale that remains finite in the
limit $a\to 0$.  It also appears in the damping rate of plasmons at
rest \cite{Bra90}:
\begin{equation}
\gamma_0 = {6.635\over 24\pi} N_c g^2T \label{e5}
\end{equation}
and in the winding number fluctuation rate at high temperature, which
is proportional to $(g^2T)^4$ \cite{Amb95}.

For quantities that depend only on $g^2T$, the physical limit
requires that the length scale $(g^2T)^{-1}$ is much larger than the
lattice spacing $a$ and much smaller than the total length of the
lattice, $Na$.  Hence, the physical limit is characterized by the two
conditions
\begin{equation}
g^2Ta \ll 1,\quad Ng^2Ta \gg 1. \label{e6}
\end{equation}
This evidently requires $N\gg 1$, which is not a surprise.  The first
condition can also be considered as a weak coupling limit; the second
one ensures that finite size effects are small.

As an explicit example, consider the plasmon damping rate $\gamma_0$,
which involves an integral over the dimensionless variable
$\xi=k/\mu$, which for $N_c=2$ has an upper limit
\begin{equation}
\xi_{\rm max} = {\pi\over a\mu} \approx {5.04\over N_c^{1/2}} 
(g^2Ta)^{-1/2}. \label{e7}
\end{equation}
On the other hand, the integration over $\xi$ becomes a discrete sum
on a finite lattice with a spacing
\begin{equation}
\Delta\xi = {\pi\over Na\mu} \approx {5.04\over N N_c^{1/2}}
(g^2\sqrt{a})^{-1/2}. \label{e8}
\end{equation}
In the case of the analytical evaluation of $\gamma_0$, the continuum 
limit requires that
\begin{equation}
g^2Ta \ll 1, \quad N^2g^2Ta \gg 1, \label{e9}
\end{equation}
which is somewhat less restrictive than (\ref{e6}).

\section{Results}

We begin the discussion of the results with the extrapolation
$(t\to\infty)$ of the maximal Lyapunov exponent.  Figure 1 shows the
time evolution of
\begin{equation}
\lambda_0(t) = {1\over t} \sum_{k=1}^{t/\tau} \ln\; s_k \label{e10}
\end{equation}
where $s_k$ is the rescaling factor obtained in scaling step $k$, and
$\tau$ is the rescaling interval.  The choice of $t^{-1/2}$ as abscissa
is motivated by the desire to be able to easily extrapolate to
$t\to\infty$, combined with the empirical finding that $\lambda_0(t)$
varies approximately linearly in this variable.  The results are for
$N=10$.

As can be seen, little extrapolation is needed for $g^2E_pa\ge 1$
(curves a--c), because $\lambda_0(t)$ rapidly approaches a constant
value.  For smaller values of $g^2E_pa$ a considerable amount of
extrapolation is required, because $\lambda_0(t)$ still systematically
decreases even at $t/a=5000$ (curves d--g).  For the lowest energy
studied here, $g^2E_pa = 0.032$ (curve h), the fluctuations in
$\lambda_0(t)$ are so large that a reliable extrapolation is quite
impossible.  It is also clear from this figure that the slope method
used in \cite{Mul92} significantly overestimates the maximal Lyapunov
exponent for $g^2E_pa <1$, since it determines $\lambda_0$ through the 
exponential growth of the distance between trajectories on the time 
scale $t/a \le 100$.  

Although the linear extrapolation in the variable $t^{-1/2}$ gives
statistically quite precise values for $\lambda_0$, except at the
smallest energies, the extrapolated values have a rather large
systematic error depending on the extrapolation scheme.  In order to
significantly reduce this error much longer runs, up to $t/a
\approx 10^5$ would be required.  Unfortunately, such runs were
impossible with our currently available computing
resources.\footnote{A $t/a=10^5$ run for a $20^3$ lattice takes
about 300 hours at 300 Mflops on a Cray T-90.}

We next turn to the finite size scaling of the Lyapunov exponents.
Figure 2a shows the time dependence of $\lambda_0(t)$ for lattices of
size $N=4,6,10$, and 20 for the energy $g^2E_pa \approx 0.125$.  The
extrapolated value of $\lambda_0$ drops by a factor 3 from $N=4$ to
$N=20$.  Figure 2b, which shows the extrapolated $\lambda_0$
versus\footnote{We have chosen this scaling because it interpolates
between the $N^{-1}$ and $N^{-2}$ dependences suggested by (\ref{e6})
and (\ref{e9}), respectively, and because it represents the scaling of
finite size fluctuations on a lattice with $N^3$ sites.} $N^{-3/2}$,
demonstrates that the size dependence is much weaker for larger
energies, such as $g^2E_pa \approx 1.8$.  Our results show
systematically that finite size effects grow as $g^2E_pa$ decreases.
This result is in accord with conditions (\ref{e6}) and (\ref{e9}).

All our results for the extrapolated Lyapunov exponents $\lambda_0$
are shown in Figure 3 on a double logarithmic scale.  It is obvious
from this figure that the deviation of $\lambda_0$ from the straight
line $\lambda_0 \approx {1\over 6}g^2E_p$ at small energies is a
finite size effect.  The value for $\lambda_0$ at $g^2E_pa = 0.125$
obtained on a $20^3$ lattice (open square) is still slightly below the
conjectured scaling line.  Note, however, that this point has a
considerable systematic uncertainty due to extrapolation (see Figure 2a).  

The published data from \cite{Mul92} are also shown in Figure 3
as the open circles.  Not unexpectedly, those values, which were
obtained on a $20^3$ lattice, clearly overestimate the correct results
for $g^2E_pa < 0.5$.  A crude estimate of this effect for $g^2E_pa =0.125$
on the basis of Figure 2a indicates that the value for $\lambda_0$
obtained by the slope method should be about a factor three higher.
This agrees quite nicely with the ratio between the old and the new
results, both for $20^3$ lattices at this energy.
In order to better resolve the results for the
region $g^2E_pa \ge 1$, we show the same data on a linear scale in
Figure 4.  The new results clearly confirm the published values in
this energy region, on which the conjectured linear scaling behavior
was based \cite{Mul92}.

\section{Conclusion}

Our new results do not lend support to the conjecture by Nielsen,
Rugh, and Rugh \cite{Nie96} that the maximal Lyapunov exponent of the
classical Yang-Mills field scales as $E^{1/4}$ at small energies.
Rather, we find that the two values of $\lambda_0$ for $g^2E_pa < 0.5$
published in \cite{Mul92} were distorted by short-time effects, i.e.
the exponential divergence of the field configurations was not
measured over a sufficiently long time.

We also have found a significant finite size effect for small
energies, as already anticipated in \cite{Bir94} (section 3.5).  Our
value of $\lambda_0$ for the smallest energy $(g^2E_pa = 0.125)$ on a
$20^3$ lattice is still consistent with the scaling law \cite{Mul92}
\begin{equation}
\lambda_0 \approx {1\over 6} g^2E_p \label{e11}
\end{equation}
within the systematic uncertainties introduced by the extrapolation
$t\to\infty$.

In summary, our results show that a precise calculation of the energy
dependence of the maximal Lyapunov of the classical SU(2) gauge theory
requires lattices with $N\ge 20$ and very long evolution times $(t/a
\gg 10^4)$.  Hopefully, such calculations will be feasible in the near
future.

\subsection*{Acknowledgements}

I gratefully acknowledge illuminating discussions with T. S. Bir\'o,
U. Heinz, C. R. Hu, S. G. Matinyan, and S. Leupold.  I thank S. E. Rugh
and H. H. Rugh for valuable comments on the manuscript.  This work was
supported in part by the U.S. Department of Energy (grant
DE-FG02-96ER40945) and by a computing grant of the North Carolina
Supercomputing Center.

\begin{figure}
\def\epsfsize#1#2{1.0#1}
\centerline{\epsfbox{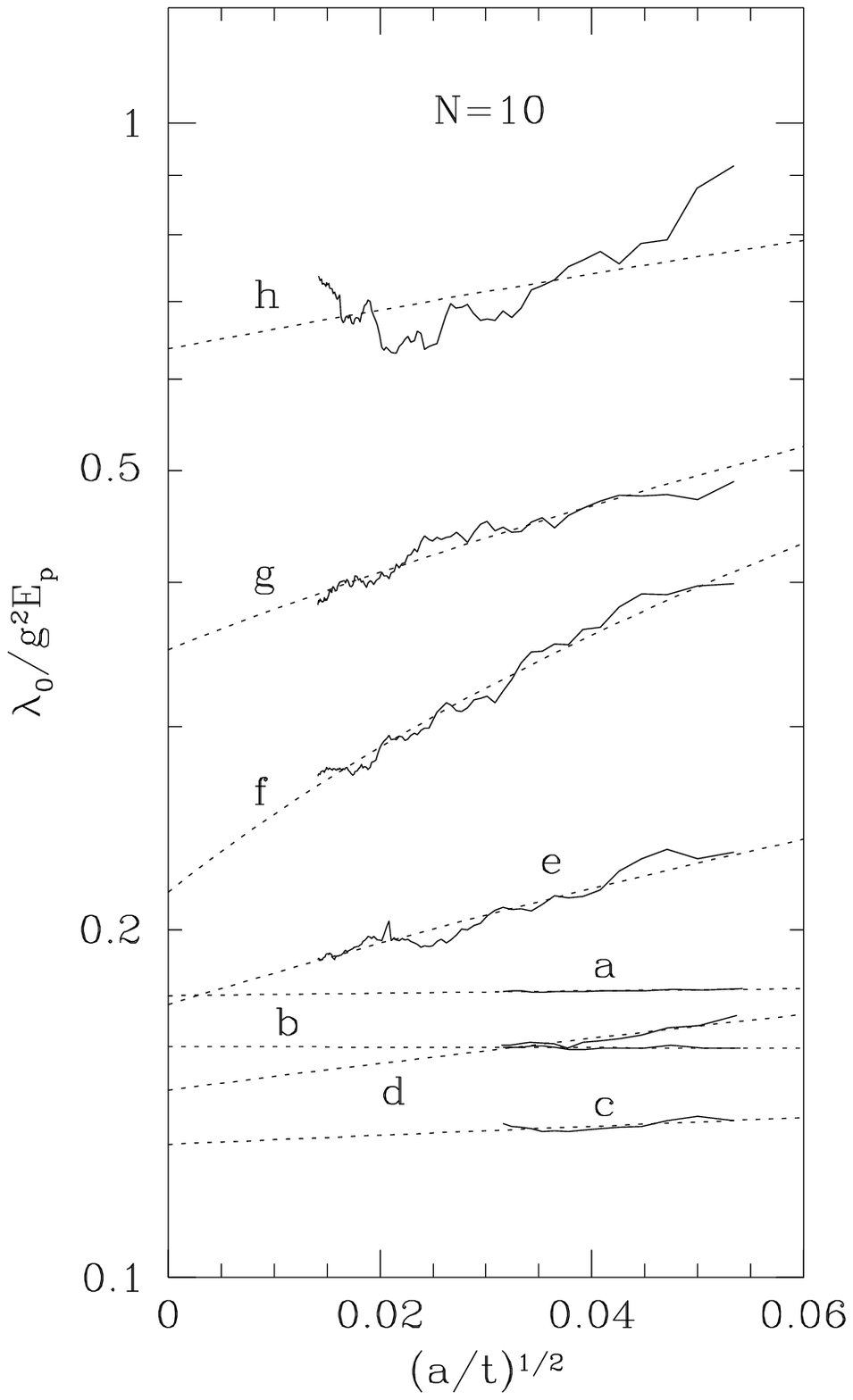}}
\caption{Time dependence of the measured Lyapunov exponent $\lambda_0$
on a $10^3$ lattice for eight energies.  The curves are labeled in
alphabetical order; the associated values of $g^2E_pa$ and the
extrapolated Lyapunov exponents are given in Table 1.  The dashed
lines represent linear least-squares fits to the curves.  (Some lines
look curved due to the logarithmic representation.)}
\label{fig1}
\end{figure}

\begin{figure}
\def\epsfsize#1#2{0.9#1}
\centerline{\epsfbox{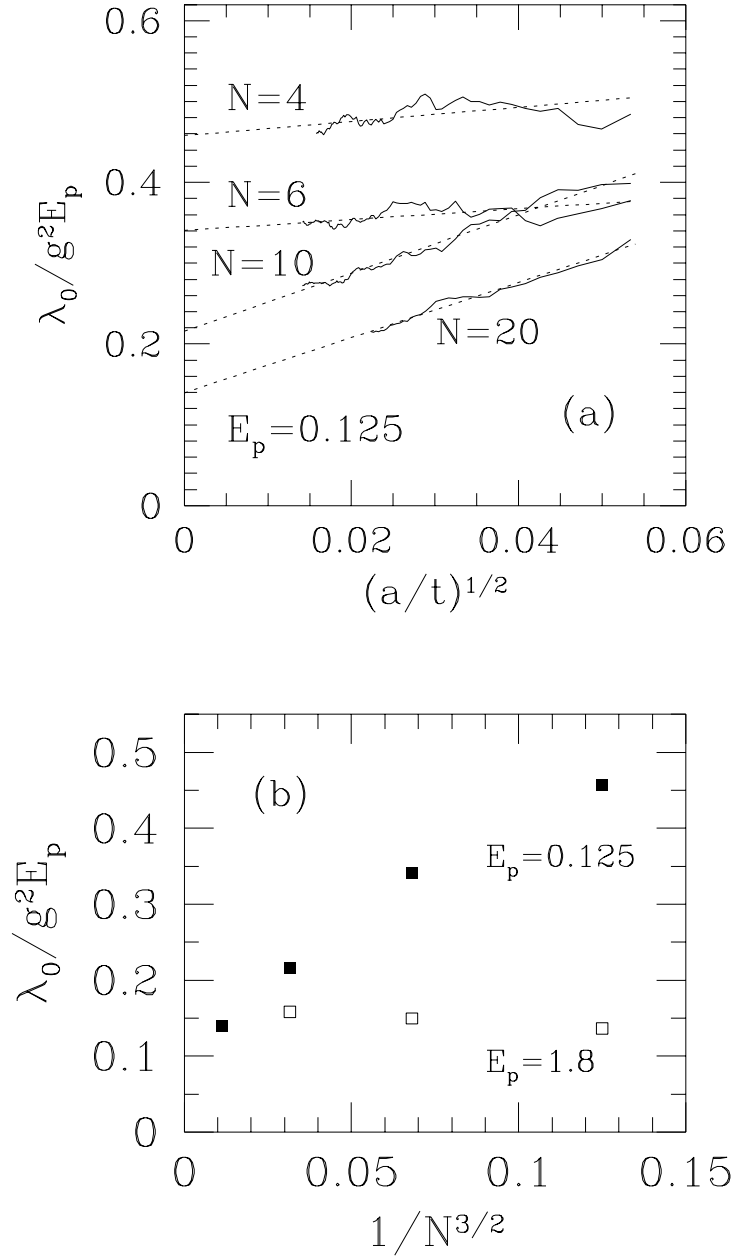}}
\caption{Finite size dependence of the Lyapunov exponents.  The upper part (a)
shows the evolution of the Lyapunov exponents obtained by the
rescaling method together with the linear extrapolation $(t\to\infty)$
for lattices of size $N=4,6,10,20$ at the energy $g^2E_pa \approx
0.125$.  The lower part (b) shows the dependence of the extrapolated
Lyapunov exponents for two different energies.}
\label{fig2}
\end{figure}

\begin{figure}
\def\epsfsize#1#2{0.9#1}
\centerline{\epsfbox{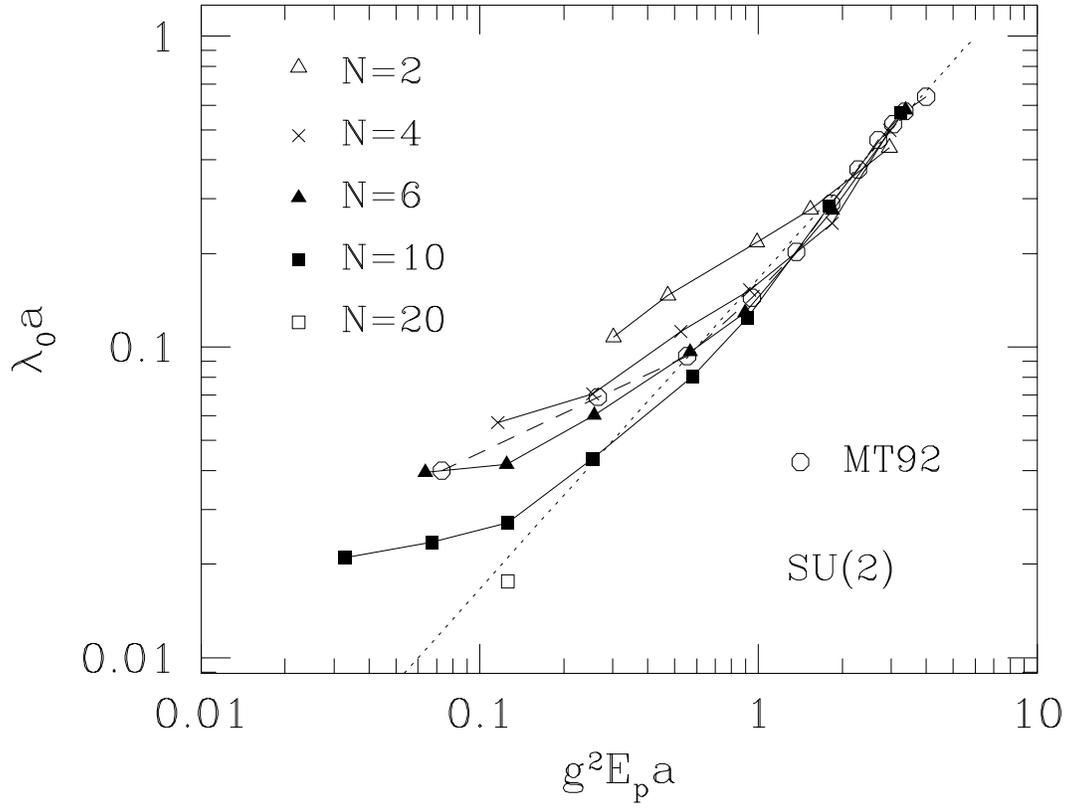}}
\caption{Energy dependence of the extrapolated Lyapunov exponents for 
lattices of size $N=2,4,6,10,20$.  The results obtained by M\"uller and
Trayanov \protect \cite{Mul92} are labelled MT92.  The dotted line is the 
fit $\lambda_0={1\over 6}g^2E_p$.}
\label{fig3}
\end{figure}

\begin{figure}
\def\epsfsize#1#2{0.9#1}
\centerline{\epsfbox{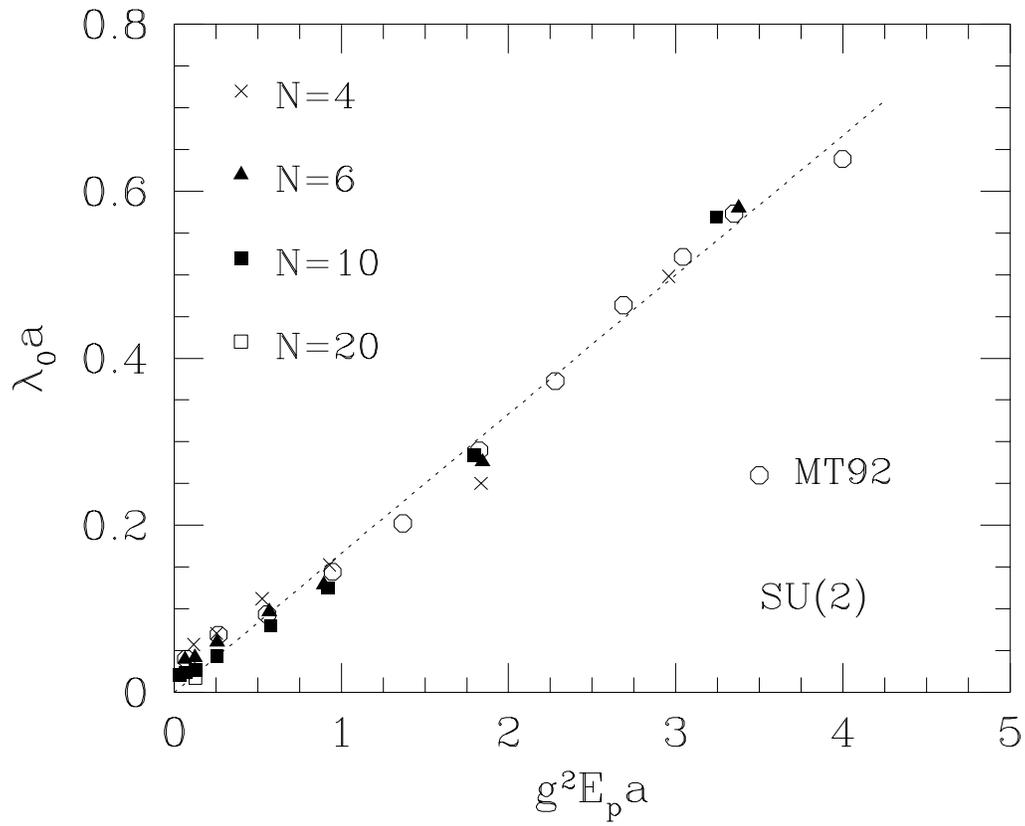}}
\caption{ Same as Figure 3, but on a linear scale.}
\label{fig4}
\end{figure}

\begin{table}
\begin{tabular}{|rcllr|} 
 $N$  & Label      & $g^2E_p a$  & $\lambda_0 a$  & $t_f/a$   \\
\hline
 4 & a       & 2.958     & 0.4981   & 1000 \\
 4 & b       & 1.836     & 0.2501   & 1000 \\
 4 & c       & 0.9288    & 0.1525   & 1000 \\
 4 & d       & 0.5265    & 0.1121   & 2000 \\
 4 & e       & 0.2542    & 0.07065  & 2000 \\
 4 & f       & 0.116     & 0.05719  & 4000 \\
\hline
 6 & a       & 3.376     & 0.5803   & 1000 \\
 6 & b       & 1.844     & 0.276    & 1000 \\
 6 & c       & 0.894     & 0.1291   & 1000 \\
 6 & d       & 0.5691    & 0.09648  & 2000 \\
 6 & e       & 0.2571    & 0.06043  & 5000 \\
 6 & f       & 0.1249    & 0.04199  & 5000 \\
 6 & g       & 0.06372   & 0.03948  & 5000 \\
\hline
10  & a      & 3.244     & 0.5691   & 1000 \\
10  & b      & 1.794     & 0.2843   & 1000 \\
10  & c      & 0.9172    & 0.1246   & 1000 \\
10  & d      & 0.5796    & 0.08039  & 1000 \\
10  & e      & 0.2535    & 0.04366  & 5000 \\
10  & f      & 0.126     & 0.02716  & 5000 \\
10  & g      & 0.06717   & 0.0235   & 5000 \\
10  & h      & 0.03291   & 0.02096  & 5000 \\
\hline
20 & f       & 0.126     & 0.0176   & 2000 

\end{tabular}

\caption{Lattice size $N$, energy per plaquette $g^2E_pa$, maximal
Lyapunov exponent $\lambda_0a$, and total evolution times $t_f/a$ for
the results presented in Figures 1--4.  The label refers to the labeling
used in Figure 1.}
\end{table}


\begin{references}

\bibitem{Amb91} J. Ambj\o rn, T. Aksgaard, H. Porter, and M.E.
Shaposhnikov, {\sl Nucl. Phys. {\bf B353}}, 346 (1991).

\bibitem{Amb92} J. Ambj\o rn and K. Farakos, {\sl Phys. Lett. {\bf
B294}}, 248 (1992).

\bibitem{Amb95} J. Ambj\o rn and H. Krasnitz, {\sl Phys. Lett. {\bf
B362}}, 91 (1995).

\bibitem{Mul92} B. M\"uller and A. Trayanov, {\sl Phys. Rev. Lett.
{\bf 68}}, 3387 (1992).

\bibitem{Gon94} C. Gong, {\sl Phys. Rev. {\bf D49}}, 2842 (1994).

\bibitem{Bir94} T.S. Bir\'o, C. Gong, B. M\"uller, and A. Trayanov,
{\sl Int. J. Mod. Phys. {\bf C5}}, 113 (1994).

\bibitem{Nie96} H.B. Nielsen, H.H. Rugh, and S.E. Rugh, ``Chaos and
Scaling in Classical Non-Abelian Gauge Fields'',
$\langle$chao-dyn/9605013$\rangle$.

\bibitem{Chi81} B.V. Chirikov and D.L. Shepelyanskii, {\sl Pis'ma Zh.
Eksp. Teor. Fiz. {\bf 34}}, 171 (1981) [{\sl JETP Letters {\bf 34}},
163 (1981)].

\bibitem{Bir95} T.S. Bir\'o, C. Gong, and B. M\"uller, {\sl Phys. Rev.
{\bf D52}}, 1260 (1995).

\bibitem{Bra90} E. Braaten and R. Pisarski, {\sl Phys. Rev. {\bf
D42}}, 2156 (1990).

\end{references}
\end{document}